\documentclass[epj]{svjour}

\usepackage{graphicx}
\usepackage{fancyhdr}
\def\makeheadbox{{
 }}
\newcommand{\U}{{\cal U}}

\def\mytitle{My title} 
\def\myauthors{My name}  
\def\mytype{My type of session}
\def\mysession{My session}

\def\mytitle{
Collider signatures for unparticle} 
\def\myauthors{
Kingman Cheung, Wai-Yee Keung, Tzu-Chiang Yuan}   
\def\mytype{Contributed Talk}    
\def\mysession{Alternatives}

\begin{document}
\title{Collider signatures for unparticle}
\author{Kingman Cheung \inst{1,2}
 \and
 Wai-Yee Keung \inst{3}
  \and
 Tzu-Chiang Yuan \inst{1,2}
}                     
%
%
\institute{Department of Physics, National Tsing Hua University,
Hsinchu, Taiwan 
\and 
Physics Division, National Center for Theoretical Sciences, Hsinchu, Taiwan
 \and
Department of Physics, University of Illinois at Chicago, Chicago IL 
60607-7059, U.S.A.
}
%
\date{}
\abstract{ 
We summarize the works presented in Refs. \cite{1,2} on collider
phenomenology of the unparticle physics associated with an exact scale
invariant sector possessing a non-trivial infrared fixed point at a
high energy scale.  We give highlights on the motivations of unparticle
and derivation of the phase space and the unparticle propagator.
We study two categories of phenomenology: (i) real
emission of unparticle that gives rise to peculiar missing energy
signatures, and (ii) virtual exchange of unparticle that interferes
with the standard model amplitudes.
} 
\maketitle
\section{Introduction}
\label{intro}
Scale invariance is a powerful concept that is widely used in various 
disciplines of physics.  However, it is manifestly broken by the masses
of elementary particles in the part of the world known to us.
Nevertheless, it is conceivable that at a much
higher scale there may exist a nontrivial
scale invariant sector with an infrared fixed point that we have not
yet probed experimentally.

Recently, Georgi \cite{georgi1} motivated by the Banks-Zaks 
(${\cal BZ}$) theory
\cite{banks-zaks} suggested that a scale invariant sector 
behaves rather peculiarly from the
perspective of particle physics. An operator $\cal O_\U$ with a general
non-integral scaling dimension $d_\U$ in a scale invariant sector has a
mass spectrum looked like a $d_{\cal U}$ number of invisible massless
particles, coined as unparticle $\U$ by Georgi.  
Unparticle does not have a fixed invariant mass but instead a 
continuous mass spectrum.
The most interesting feature is that real production of unparticle,
described by an effective field theory, can give rise to peculiar
missing energy distributions because of the possible non-integral values of
$d_\U$.

Shortly after Georgi's work \cite{georgi1}, 
the propagator for the unparticle was worked out independently 
in \cite{georgi2} and  \cite{1}.  An unusual
phase in the unparticle propagator was discovered by both groups and
the interesting interference patterns between the amplitude of 
$s$-channel unparticle 
exchange and those from the SM were studied.  
In this proceedings, we summarize the collider phenomenology presented 
in Refs. \cite{1,2}, while there have been a large amount of recent literature
on this subject but we have to skip because of space.

\section{Formalism}
\label{sec1}
Effective field theory approach was adopted in studying the phenomenology
of unparticle \cite{georgi1}.  The $\cal BZ$ sector interacts with the
SM fields through the exchange of a connector sector at 
a high mass scale $M_\U$. Below $M_\U$ non-renormalizable operators 
suppressed by inverse powers of
$M_\U$ are induced. Generically, operators are of the form
\begin{equation}
   \frac{ 1 } {M_\U^{d_{\mathrm SM} + d_{\cal BZ}-4}} \, 
   {\cal O}_{\mathrm SM} \, {\cal O}_{\cal BZ}  \; ,
\label{genericop}
\end{equation}
where ${\cal O}_{\mathrm SM}$ and ${\cal O_{BZ}}$ represent local operators
constructed out of SM and ${\cal BZ}$ fields 
with scaling dimensions $d_{\mathrm SM}$ and $d_{\cal BZ}$,
respectively. 
Renormalization effects in the scale invariant $\cal BZ$ sector induce
dimensional transmutation \cite{coleman-weinberg} at an energy scale
$\Lambda_\U$ . Below $\Lambda_\U$ matching conditions must be imposed
onto the operator (\ref{genericop}) to match a new set of operators
having the form
\begin{equation}
 \label{effectiveop}
    C_{\cal O_U} \frac{ \Lambda_\U^{d_{\cal BZ} - d_\U} } {M^{d_{\mathrm SM} + d_{\cal BZ}-4}_\U } \,
   {\cal O}_{\mathrm SM}\, {\cal O}_\U \;,
\end{equation}
where  $d_\U$ is the scaling dimension of the unparticle operator 
${\cal O_U}$ and
$C_{\cal O_\U}$ is a coefficient function fixed by the matching.
We assume that exact scale invariance survives all the way down 
to the electroweak scale.

It was demonstrated in \cite{georgi1} that  scale invariance can be 
used to fix the two-point functions of 
the unparticle operators. 
The spectral density $\rho_\U(P^2)$ is then given by
\begin{eqnarray}
\rho_\U(P^2)
&=& \int d^4 x  \,  e^{i P \cdot x}\langle 0| O_\U(x)
   O_\U^\dagger(0)|0\rangle \nonumber \\
   & = & A_{d_\U} \ \theta(P^0) \ \theta(P^2) \ (P^2)^\alpha
\end{eqnarray}
where $\alpha$ is fixed by scale invariance to be $(d_\U-2)$,
and $A_{d_\U}$ is normalized to interpolate 
the $d_\U$-body phase space of  massless
particle \cite{georgi1}, given by
\begin{equation}
   A_{d_\U}={16\pi^2\sqrt{\pi}\over (2\pi)^{2{d_\U}}}
       { \Gamma({d_\U}+{1\over
       2})\over\Gamma({d_\U}-1)\Gamma(2\,{d_\U})}  \; .
\end{equation}
The most important feature of unparticle is 
that $d_\U$ can take on non-integral values, and so
we can imagine something similar to fractional particles.

\subsection{Unparticle propagators}
The derivation of the virtual unparticle propagator is also based on
scale invariance.  Without loss of generality we consider a 
scalar propagator. The extensions to spin-1 and spin-2 propagators
simply include the appropriate spin structures.
The Feynman propagator $\Delta_{F} (P^2)$ of the unparticle is determined 
by the spectral formula
\begin{eqnarray}
\Delta_{F} (P^2)&=&
 \frac{1}{2\pi} - \!\!\!\!\!\!\! \int_0^\infty \frac{R(M^2) dM^2}{P^2-M^2} 
- i \frac{1}{2} R(P^2) \theta(P^2)
\end{eqnarray}
where $R(M^2) = A_{d_\U} (M^2)^{d_\U -2}$.
The appropriate form for $\Delta_{F} (P^2)$ 
is $\Delta_{F} (P^2) = Z_{d_\U} (-P^2)^{d_\U -2}$, 
where $Z_{d_\U}$ is the factor to be determined.
The complex function $(-P^2)^{d_\U -2}$ is analytic for negative $P^2$,
but needs a branch cut for positive $P^2$:
\begin{equation}
\label{branchcut}
 (-P^2)^{d_\U -2}=\left \{
\begin{array}{lcl}
|P^2|^{d_\U -2}   & \quad & \hbox{if $P^2 \le 0 $, } \\
|P^2|^{d_\U -2} e^{-i d_\U \pi} & & \hbox{if $P^2 > 0$.} 
\end{array} \right.
\end{equation}
The factor $Z_{d_\U}$ is determined by comparing 
with the imaginary part of $\Delta_F (P^2)$ for a 
time-like momentum $(P^2>0)$ and given by
\begin{equation}
  Z_{d_\U}  = \frac{A_{d_\U}} { 2 \sin (d_\U \pi) } \;.
\end{equation}

\subsection{Effective interactions}
A few common 
effective interactions that satisfy the standard model gauge symmetry 
for the scalar, vector, and tensor unparticle operators with SM 
fields are given, respectively, by
\begin{eqnarray}
&&
\lambda_0 \frac{ 1}{\Lambda_\U^{d_\U-1}} \bar f f O_\U\;, \;
\lambda_0 \frac{1}{\Lambda_\U^{d_\U} } G_{\alpha\beta} G^{\alpha\beta}
O_\U \;, 
\label{lambda0} \\
&&\lambda_1 \frac{1}{\Lambda_\U^{d_\U - 1} }\, \bar f \gamma_\mu f \,
O_\U^\mu \;, \;\;
\lambda_1 \frac{1}{\Lambda_\U^{d_\U - 1} }\, \bar f \gamma_\mu \gamma_5 f \,
O_\U^\mu \;, 
\label{lambda1} \\
&&- \frac{1}{4}\lambda_2 \frac{1}{\Lambda_\U^{d_\U} } \bar \psi \,i
   \left(  \gamma_\mu \stackrel{\leftrightarrow}{D}_\nu + 
           \gamma_\nu \stackrel{\leftrightarrow}{D}_\mu \right )
  \psi  \,  O_\U^{\mu\nu} \;, \nonumber \\
&&
   \lambda_2 \frac{1}{\Lambda_\U^{d_\U} } G_{\mu\alpha}
G_{\nu}^{\;\alpha} O_\U^{\mu\nu} 
 \;, 
\label{lambda2}
\end{eqnarray}
where the covariant derivative $D_\mu = \partial_\mu + i g \frac{\tau^a}{2} 
W^a_\mu + i g' \frac{Y}{2} B_\mu$, 
$G^{\alpha\beta}$ denotes the gauge field strength (gluon, photon and 
weak gauge bosons), and $\lambda_{i}$ are dimensionless effective
couplings $C_{O^i_\U} \Lambda_\U^{d_{\cal BZ}}/M_\U^{d_{\rm SM} + 
d_{\cal BZ}-4}$.

Virtual exchange of unparticle corresponding to the vector operator
$O_\U^\mu$ between two fermionic currents
can result in 4-fermion interactions \cite{1,2,georgi2}
\begin{eqnarray}
{\cal M}_1^{4f} = \lambda_1^2 \, Z_{d_\U} 
\frac{1}{\Lambda_\U^2}
 \left(
- \frac{P_\U^2}
{\Lambda_\U^2} 
  \right)^{d_\U - 2} \, 
\bar f_2 \gamma_\mu f_1\,  \bar f_4 \gamma^\mu f_3\;.
\label{4fermionsop}
\end{eqnarray}
The phase factor $\exp (-i\pi d_{\cal U})$ for time-like
momentum  $P_{\cal U}^2>0$ can give rise to nontrivial interference
patterns with SM amplitudes.
Another important feature is that the high
energy
behavior of the amplitude scales as $(s/\Lambda_\U^2 )^{d_\U  -1}$.  
For $d_\U=1$ the tree amplitude behaves like that of a massless
photon exchange, 
while for $d_\U = 2$ the amplitude reduces to the
conventional 4-fermion contact interaction \cite{contact}.
If $d_\U$ is between 1 and 2, say $3/2$, the
amplitude has the unusual behavior of $\sqrt{s}/\Lambda_\U$ at
high energy.  
If $d_\U = 3$ the amplitude's high energy behavior becomes $(
s/\Lambda_\U^2)^2$,
which resembles that of Kaluza-Klein tower of gravitons.
One can also consider a spin-2 unparticle exchange between a pair of
fermionic currents, but 
it is suppressed by $(s/\Lambda_\U)^2$
relative to that induced by spin-1 unparticle operator.  

\section{Collider Phenomenology}
\label{sec2}

\subsection{Real emissions}
\begin{figure}[t!]
\includegraphics[width=3.3in]{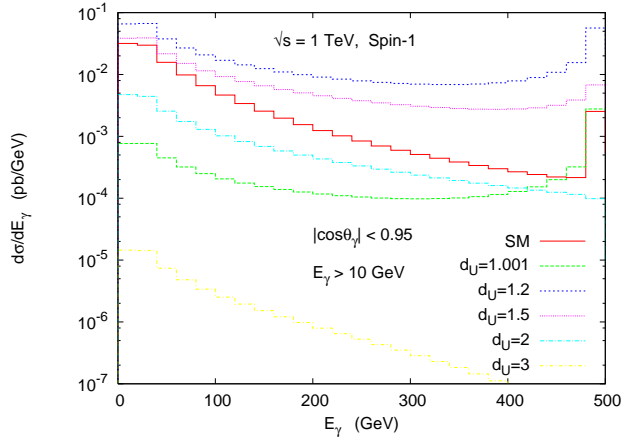}
\caption{
\label{eegammau} \small
Comparison of photon energy spectrum
of $e^- e^+ \to \gamma {\cal U}$ (spin-1 unparticle) with 
the SM background $e^-e^+\to \gamma Z^* \to \gamma \nu \bar \nu$ 
for different values of $d_{\cal U}=1.001,\,1.2,\,1.5,\,2$ and 3 at 
$\sqrt s = 1$ TeV.
}
\end{figure}

{\it Mono-photon events in $e^- e^+$ collisions}: The energy
spectrum of
the mono-photon from the process $e^- e^+ \to \gamma \U $ 
can be used to probe the unparticle sector.  Its
cross section is given by
\begin{equation} 
 d \sigma = \frac{1}{2 s} \,  | \overline{{\cal M}} |^2 \;
 \frac{  A_{d_\U} }{ 16 \pi^3 \Lambda_\U^2} 
 \left( {P^2_\U\over \Lambda_{\cal U}^2} \right )^{ d_\U - 2}  
\, E_\gamma d E_\gamma d \Omega  
\end{equation}
with the matrix element squared 
\begin{equation}
  |\overline{{\cal M}}|^2 = 2 e^2 Q_e^2 \lambda_1^2 \,
  \frac{ u^2 + t^2 + 2 s P^2_\U}{ u t} \; .
  \end{equation}
The $P^2_\U$ is related to the energy of the photon $E_\gamma$ by the
recoil mass relation, 
$  P^2_\U = s - 2 \sqrt{s} \, E_\gamma$.
The mono-photon energy spectrum is
plotted in Fig.~\ref{eegammau} for various choices of $d_\U$. The
sensitivity of the scaling dimension to the energy distribution can be
easily discerned.  

\begin{figure}[t!]
\includegraphics[width=3.3in]{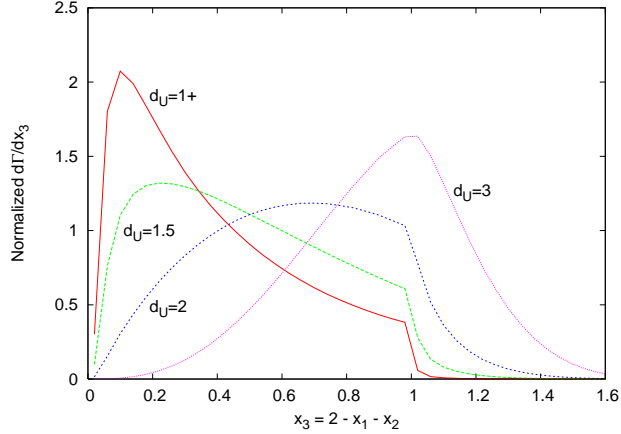}
\caption{
\label{zqqu} \small
Normalized decay rate of $Z \to q \bar q {\cal U}$  for spin-1 unparticle 
versus $x_3 = 2 - x_1 - x_2$ for different values of 
$d_{\cal U}=1^+,\,1.5,\,2$, and 3, where ``$1^+$'' stands for $1+\epsilon$
for a small positive $\epsilon$.
}
\end{figure}

{$Z \to f \bar f \U$}: The decay width for the process can be
easily obtained as 
\begin{eqnarray}
\label{ztoqqU}
\frac{1}{\Gamma_{Z \to f \bar f} }
\frac{d\Gamma(Z \to f \bar f +{\cal U}) }
{ d x_1 d x_2 d \xi} & =
 &
 \frac{\lambda_1^2}{8\pi^3} \, g(1-x_1,1-x_2,\xi) \nonumber \\
& \times & \frac{M_Z^2} {\Lambda_\U^2}
 A_{d_{\cal U}} \left(\frac{P_\U^2}{\Lambda_\U^2}\right)^{d_{\cal
 U}-2}
\end{eqnarray}
where $ \xi = P_\U^2/M_Z^2$ and $x_{1,2}$ are the energy fractions of
the fermions $x_{1,2} = 2 E_{f, \bar f} / M_Z$, and the function
$g(z,w,\xi)$ is given in Ref. \cite{1}.
In Fig.~\ref{zqqu}, we plot the normalized decay rate of this
process versus the energy fraction $x_3=2-x_1 - x_2$.  One can see
that the shape depends sensitively on the scaling dimension of the
unparticle operator. 

\subsection{Virtual Exchanges}

\begin{figure}[t!]
\includegraphics[width=3.3in]{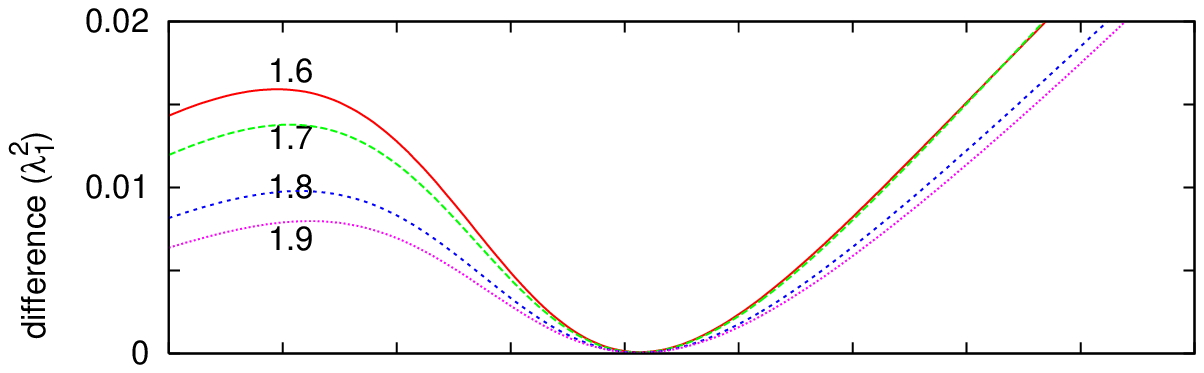}

\vspace{-0.29in}

\includegraphics[width=3.3in]{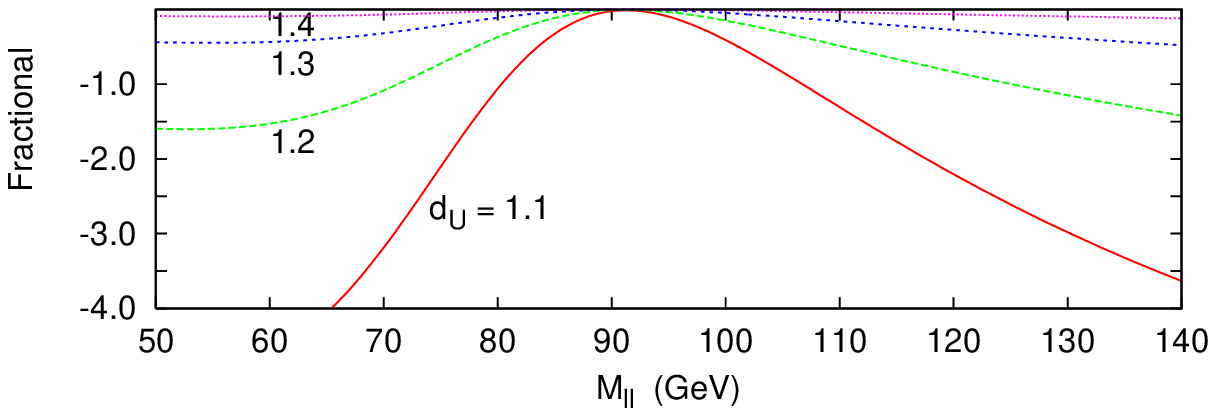}
\caption{
\label{drell-yan} \small
Fractional difference from the SM prediction of the Drell-Yan
invariant mass spectrum for various $d_{\cal U}$ at the Tevatron in
units of $\lambda_1^2$.  We have chosen $\Lambda_\U = 1$ TeV. 
The curve for 
$d_\U=1.5$ is too close to zero for visibility in the current scale.
}
\end{figure}

{\it Drell-Yan} production including the effect of the spin-1 unparticle 
was first studied in Ref. \cite{1}. 
The differential cross section is given by
\begin{eqnarray}
\frac{d^2 \sigma}{ d M_{\ell\ell} \, dy}  &=&  K
  \frac{M^3_{\ell\ell}}{72\pi s}\,
 \sum_q \, f_q(x_1) f_{\bar q}(x_2) \;
 \times \biggr (
  | M_{LL} |^2  \nonumber \\
&& +   | M_{LR} |^2  +  | M_{RL} |^2  +  | M_{RR} |^2 
             \biggr ) \,,
\end{eqnarray}
where $\hat s = M^2_{\ell\ell}$ and $\sqrt{s}$ is the center-of-mass
energy of the colliding hadrons.  $M_{\ell\ell}$ and $y$ are
the invariant mass and the rapidity of the lepton pair, respectively,
and $x_{1,2} = M_{\ell\ell}e^{\pm y}/\sqrt{s}$.
The $K$ factor equals $1 + \frac{\alpha_s}{2\pi} \frac{4}{3} \left(
  1+ \frac{4 \pi^2}{3} \right )$.
The reduced amplitude $M_{\alpha\beta} (\alpha,\beta = L,R)$ is given
by
\begin{eqnarray}
 M_{\alpha \beta}  &=&  
  \     \lambda_1^2 Z_{d_\U}  \frac{1}{\Lambda_\U^2} 
 \left (- \frac{\hat s}{\Lambda_\U^2} \right)^{d_\U-2} 
+ \frac{ e^2 Q_l Q_q}{ \hat s} \nonumber \\
&&
  + \frac{e^2 g^l_\alpha g^q_\beta}{ \sin^2 \theta_{\rm w}
  \cos^2\theta_{\rm w}
 }\, \frac{1}{\hat s - M_Z^2+iM_Z\Gamma_Z} \; .
\label{reduced}
\end{eqnarray}
This unparticle propagator with the phase factor \\
$\exp(-i \pi d_\U )$
interferes with both the real photon propagator and 
the real and imaginary parts of the unstable $Z$ boson propagator.  
This gives rise to interesting interference patterns \cite{georgi2,1}.
Effects from spin-2 unparticle exchanges are suppressed w.r.t. spin-1 exchanges
\cite{2}. 
In Fig.~\ref{drell-yan}, we depict the 
fractional difference from the SM prediction 
in units of $\lambda_1^2$ (with small $\lambda_1$ 
while keeping $\Lambda_\U = 1$ TeV) of the Drell-Yan distribution as
a function of the invariant mass of the lepton pair for various 
$d_\U$. 

{$e^+ e^- \to f \bar f$} 
production at $e^- e^+$ colliders can be studied using
the amplitude in Eq.~(\ref{reduced}) 
with appropriate color-factor modifications
for spin-1 unparticle exchange.
The differential cross section including the spin-1 unparticle 
exchange is given by
\begin{eqnarray}
\frac{d\sigma (e^- e^+ \to f\bar f)}{d \cos\theta} &=&
  \frac{N_c s}{128 \pi} \biggr[ ( 1+ \cos\theta)^2 ( |M_{LL}|^2 \nonumber \\
 && \hspace{-1in}
+ |M_{RR}|^2 )
                             + ( 1- \cos\theta)^2 ( |M_{LR}|^2 + |M_{RL}|^2 ) 
                        \biggr ] \;.
\label{ffeespin1}
\end{eqnarray}
The unparticle 4-fermion contact interactions in 
Eq. (\ref{4fermionsop})
can be different for different chiralities of the fermions. 
It is clear from Eq.~(\ref{ffeespin1}) that different modifications to 
$M_{\alpha\beta}$ can significantly change the angular distribution, because 
$M_{LL}$ and $M_{RR}$ are multiplied by $(1+\cos\theta)^2$ while 
$M_{LR}$ and $M_{RL}$ are multiplied by $(1-\cos\theta)^2$.
The forward-backward asymmetry can therefore discriminate
various chirality couplings.

\begin{figure}[t!]
\includegraphics[width=3.2in]{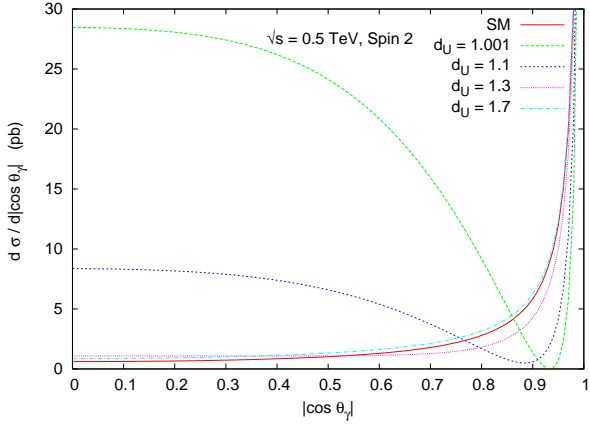}
\caption{
\label{eegg-spin2-lambda5} \small
The differential cross section 
$\frac{d \sigma}{d |\cos\theta_\gamma|} ( e^- e^+ \to \gamma\gamma)$
versus $|\cos\theta_\gamma|$ at $\sqrt{s} = 0.5$ TeV with a spin-2 unparticle
exchange plus SM contributions.  $\lambda_2$ is 
set at $5$ for visibility and $\Lambda_{\U} =1 $ TeV.}
\end{figure}

{\it Diphoton production}
at $e^- e^+$ and hadronic colliders is 
very useful to detect unknown resonances that decay into a pair of
photons and to search for anomalous diphoton couplings.  The spin-2 unparticle
can couple to a pair of fermions via the first operator of Eq.~(\ref{lambda2}) 
and to a pair of photons via the second operator in Eq.~(\ref{lambda2}).
The amplitudes are given in Ref. \cite{2}.
We show the angular distribution in Fig. \ref{eegg-spin2-lambda5}.  In the SM,
the angular distribution is very forward with majority of the cross section
at $|\cos\theta_\gamma|$ close to 1.  When $d_U$ is less than $1.2$ the 
majority comes  from the central region and a dip is formed around 
$|\cos\theta_\gamma| \approx 0.9$.  It is because of the spin-2 structure
of the operator.

\subsection{Present constraints}
\begin{figure}
\centering
\includegraphics[width=3.2in]{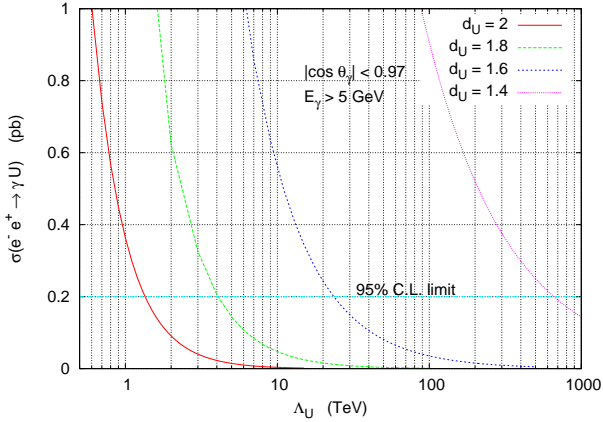}
\caption{\small \label{eegU-limit}
Cross sections for mono-photon plus unparticle production at 
the $e^- e^+$ collider with $\sqrt{s} = 207$ GeV for 
$d_\U =$ 1.4, 1,6, 1.8 and  2. 
The horizontal line of 0.2 pb is the 95\% C.L. upper limit.
}
\end{figure}

The best limit on single-photon production came from 
L3 \cite{lep-ph}, which obtained an 95\% C.L.
upper limit on $\sigma ( e^- e^+ \to \gamma + X) \simeq 0.2$ pb under
the cuts: $E_\gamma > 5$ GeV and $|\cos \theta_\gamma |< 0.97$ at
$\sqrt{s} = 207$ GeV.
We calculate mono-photon plus unparticle production with the same cuts
in $e^- e^+$ collisions with $\sqrt{s}=207$ GeV versus the unparticle
scale $\Lambda_\U$ (with a fixed $\lambda_1 = 1$) for 
$d_\U = $ 1.4, 1.6, 1.8 and 2 in 
Fig. \ref{eegU-limit}.  We have also drawn the horizontal line showing
the 95\% C.L. upper limit (0.2 pb).  
Since the production cross section scales as 
$\lambda_1^2/ \Lambda_\U^{2d_\U - 2}$, the limits increases very rapidly
when $d_U$ decreases from 2 to 1.4 with $\lambda_1$ fixed.  

\begin{figure}[t!]
\centering
\includegraphics[width=3.2in]{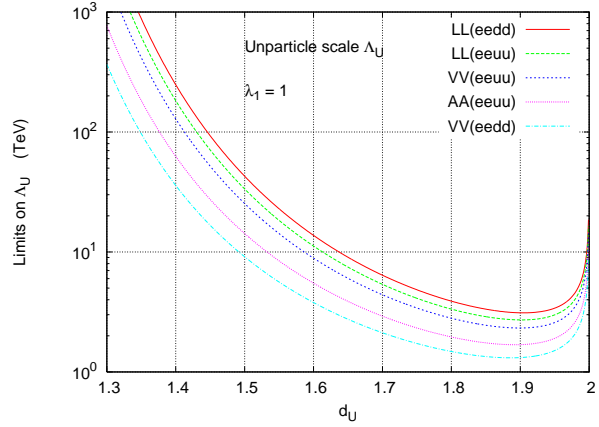}
\caption{ \small \label{limit-4f}
Rescaled limits from  existing 4-fermion contact interactions.  $LL$ means
only left-left chirality is considered while $VV$ means $LL+RR+LR+RL$
and $AA$ means $LL+RR-LR-RL$.  We have chosen $P_\U^2 \approx 
(0.2\; {\rm TeV} )^2$.
}
\end{figure}

Since spin-1 unparticle exchanges will lead to 4-fermion contact
interactions, we can use the existing limits on 4-fermion contact
interactions \cite{contact-data,pdg} to constrain the unparticle scale
$\Lambda_\U$.  We can compare Eq.~(\ref{4fermionsop}) with the conventional 
4-fermion contact interactions
\begin{equation}
{\cal L}_{4f} = \frac{4 \pi}{\Lambda^2} 
  \sum_{\alpha, \beta =L,R} \, \eta_{\alpha\beta}
( \bar e \gamma_\mu P_\alpha e)\,  ( \bar f \gamma^\mu P_\beta f) \;,
\end{equation}
which results in the following equality:
\begin{equation}
\lambda_1^2 \, Z_{d_\U} \,\frac{1}{\Lambda_\U^2}
 \,  \left( - \frac{P_\U^2}{\Lambda_\U^2}  \right)^{d_\U - 2} 
  =  \frac{4 \pi}{\left( \Lambda^{95} \right)^2} \;,
\label{est}
\end{equation}
where $\Lambda^{95}$s are the 95\% C.L. limits on the $eeqq$ 
contact interaction scales obtained by combining global data 
\cite{contact-data}.
The best limit is on the $LL$ chirality because the parity-violating
experiments, especially the atomic-parity violation, are very stringent:
$\Lambda^{95}_{LL}(eeuu)$ $\simeq  23 $ TeV while 
$\Lambda^{95}_{LL}(eedd) \simeq  26 $ TeV.  
The results are shown in Fig. \ref{limit-4f}.  

Phenomenology of unparticle is very rich.
While the underlying theory of unparticle is still needed to be unraveled by
theorists, experimentalists could detect such a hidden scale invariant
sector when the LHC turns on in 2008.

The work was supported in part by the NSC of Taiwan 
and U.S. DOE.

\end{document}